\documentclass[acmsmall]{acmart}

\usepackage{amsmath}
\usepackage{algorithmic}
\usepackage{multirow}
\usepackage{tcolorbox}
\usepackage{IEEEtrantools}

\setcopyright{acmlicensed}
\copyrightyear{2018}
\acmYear{2018}
\acmDOI{XXXXXXX.XXXXXXX}

\acmJournal{JACM}
\acmVolume{37}
\acmNumber{4}
\acmArticle{111}
\acmMonth{8}

\begin{document}

\title[Are LLM-Generated Formal Specifications Faithful to Their Requirements?]
{Are LLM-Generated Formal Specifications Faithful to Their Requirements? A Study on Bimodal Formal Specification Generation}

\author{Wenhan Wang}
\email{wangwenhan@isacas.ac.cn}
\affiliation{%
  \institution{Institute for Software Chinese Academy of Sciences}
  \country{China}
}

\author{Shiming Chen}
\email{csm76456@163.com}
\affiliation{%
  \institution{Beijing Jiaotong University}
  \country{China}
}

\author{Zeyu Sun}
\email{zeyu.zys@gmail.com}
\affiliation{%
  \institution{Institute for Software Chinese Academy of Sciences}
  \country{China}
}

\author{Shenghua Feng}
\email{fengshenghua@isacas.ac.cn}
\affiliation{%
  \institution{Institute for Software Chinese Academy of Sciences}
  \country{China}
}

\begin{abstract}
Formal specification generation is a critical step towards formal verification and correctness assurance of software. Recently, large language models have been widely adopted in automatic generation of formal specifications.
However, existing LLM-based specification generation approaches often assume that the input program is correct, which deviates from the real purpose of formal verification---to figure out whether the program meets its requirements when its correctness is unknown.
Therefore, we conduct an empirical study of LLM-based specification generation under the bimodal setting: given both the natural language requirement and its corresponding code implementation, whose correctness is unknown. We construct a benchmark of 1,403 C programs and evaluate ACSL specification generation with Frama-C verification.
Our study reaches several interesting findings: first, when the code implementation is correct, providing the natural language requirement does not further improve the pass rate of generated specifications. Second, when the code implementation is incorrect, LLM-based specification generation may erroneously pass verification by producing weak or implementation-misled specifications, and this cannot be fully alleviated by advanced prompting approaches. Compared with the non-formal HoarePrompt approach, LLM-generated formal specifications are less effective in predicting requirement-code consistency.
Our study takes a first look at applying LLM-based formal specification generation to verifying program correctness against requirements, and points to requirement-faithful specification generation as a new direction for future LLM-based formal specification research.

\end{abstract}


\maketitle

\section{Introduction}
Formal specifications connect natural language requirements with machine-checked program verification. In specification languages such as ACSL and JML, these specifications are not merely documentation: they define the properties that a verifier checks against an implementation. Recent advances in large language models (LLMs) make it increasingly plausible to generate such specifications automatically. However, for LLM-generated specifications to support verification in practice, they must satisfy two requirements at the same time: they must capture the intended behavior described by the requirement, and they must contain enough implementation-aware detail, such as memory constraints, control-flow conditions, and loop invariants, to be verified by formal tools.

However, existing studies have not fully addressed this bimodal nature of formal specification generation. Code-based specification generation can exploit implementation structure and often produces specifications that are easier to check, but it commonly assumes that the implementation is correct and treats verifier success as a positive outcome. This assumption weakens the connection to the original requirement: if the implementation is wrong, a specification inferred from the implementation alone will simply encode the wrong behavior. Requirement-based generation, in contrast, is closer to the intended behavior, but without the implementation it is difficult to infer proof-relevant specification components needed by verifiers such as loop invariants and assertions other than pre- and post-conditions. Some recent works have explored LLM-based specification generation for defect prediction against natural language requirements, but they either generate simple non-verifiable assertions~\cite{endres2024can}, or generate specifications from correct programs and use them to detect bugs in mutants~\cite{richter2025beyond}. These settings do not directly answer whether LLMs can generate formal specifications from both natural language requirements and code implementations whose correctness is unknown.

This missing setting creates a faithfulness gap. In program verification, the semantic oracle should be the requirement, not the implementation. For a correct implementation, a requirement-faithful specification should pass verification. For an incorrect implementation, however, a requirement-faithful specification should fail verification because it exposes the mismatch between the expected behavior and the actual behavior. Therefore, in bimodal specification generation, verification pass is not always a success signal. A specification that passes on an incorrect implementation may be too weak, or it may have been misled by the implementation and captured the wrong behavior. Understanding this gap is essential before LLM-generated specifications can be used as reliable evidence for program correctness.

In this paper, we conduct an empirical study of LLM-based formal specification generation from bimodal sources. We construct a benchmark of 1,403 programs from 500 programming-contest problems, covering both human-written correct and incorrect implementations. Each benchmark instance contains a natural language requirement and a normalized implementation that can be checked with ACSL and Frama-C. Our study asks whether LLMs can generate specifications that are not only syntactically accepted and verifiable, but also faithful to the requirement when the implementation may be wrong.

In our study, we aim to answer the following research questions:

\begin{itemize}
    \item \textbf{RQ1: How do bimodal information sources affect the correctness of LLM-generated formal specifications for program verification?}
    For correct implementations, we compare requirement, code, and combined inputs to study which modality better supports verifiable ACSL generation.

    \item \textbf{RQ2: When the code implementation is incorrect, can LLMs generate formal specifications that are consistent with the requirements?}
    This RQ addresses the central faithfulness question: generated specifications should capture the \textbf{expected program behavior} in the requirement rather than the \textbf{actual behavior} of the implementation. We therefore analyze verification outcomes and classify specifications as correct, wrong, weak, or misled.

    \item \textbf{RQ3: Can advanced prompting and workflow design improve LLM performance on bimodal specification generation?}
    We study whether task decomposition and iterative refinement can mitigate the faithfulness gap by separating requirement reasoning from implementation- and verifier-guided specification refinement.

    \item \textbf{RQ4: How does LLM-based bimodal specification generation compare with non-formal approaches in predicting the coherence between requirements and implementations?}
    We compare our formal specification-based pipeline with HoarePrompt~\cite{bouras2025hoareprompt}, a semi-formal LLM-based approach for requirement-code consistency reasoning.
\end{itemize}

Our results reveal several important challenges for LLM-based formal specification generation. When the implementation is correct, providing the natural language requirement does not necessarily improve the verification rate of generated specifications. More importantly, when the implementation is incorrect, LLM-generated specifications may pass verification for the wrong reason: they may be too weak to distinguish the bug, or they may overfit to the incorrect implementation. These findings suggest that future LLM-based specification generators should be evaluated not only by syntactic correctness or verifier pass rates, but also by requirement faithfulness.

In summary, our contributions are as follows:
\begin{itemize}
    \item We conduct an empirical study of LLM-based formal specification generation from bimodal sources, where both natural language requirements and code implementations are provided without assuming implementation correctness.
    
    \item We construct a benchmark of 1,403 programs from 500 problems, including both correct and incorrect human-written implementations for requirement-level program verification.
    
    \item We analyze the faithfulness of LLM-generated ACSL specifications and study whether advanced prompting and refinement workflows can reduce weak or implementation-misled specifications.
\end{itemize}

Overall, our study shows that bimodal formal specification generation is not only a generation problem, but also a requirement-faithfulness problem. These findings provide empirical evidence and evaluation guidance for future research on LLM-driven specification generation and program verification.

\section{Related Work}
Recently, LLMs have been frequently adopted in formal specification generation from code or natural language for program verification, surpassing traditional approaches that rely on static analysis or pre-defined templates. 

A large amount of existing works focus on generating specifications from source code~\cite{pei2023can,wen2024enchanting,ma2026specsyn,ma2025specgen,le2025can,chen2025automated,yang2025autoverus,wang2026kbspec}. Pei et al.~\cite{pei2023can} fine-tuned an LLM for generating Java program invariants. AutoSpec~\cite{wen2024enchanting} and SpecSyn~\cite{ma2026specsyn} integrate LLMs and static analysis in generating ACSL specifications. SpecGen~\cite{ma2025specgen} leverages iterative repair from verifier feedback and a random mutation mechanism for LLM-based JML specification generation. Le-Cong et al.~\cite{le2025can} proposed a large-scale benchmark, FormalBench, for evaluating LLMs on JML specification generation. Some other works~\cite{chen2025automated,yang2025autoverus} target Rust and generate both Verus specifications and proof.

Some other works target specification generation from natural language.
The majority of works in this line focus on temporal logic specifications~\cite{cosler2023nl2spec,fang2025enhancing,chen2023nl2tl}, instead of specification languages that support program verification. Endres et al.~\cite{endres2024can} leverage LLMs for generating post-conditions from natural language requirements. Cao et al.~\cite{cao2025informal} performed an empirical study on multiple natural language to specification tasks across 5 specification languages. Yang et al.~\cite{yang2026how} conducted a comprehensive study on the capabilities of LLMs in generating Rocq formal specifications from HumanEval task descriptions.
Some works adopt LLMs in simultaneously generating code and specification for Dafny~\cite{misu2024towards,sun2024clover,banerjee2026dafnypro,xu2025local}, a programming language that officially supports formal specification and verification.

\section{Methodology}

\begin{figure*}[t]
\centering
\includegraphics[width=0.8\textwidth]{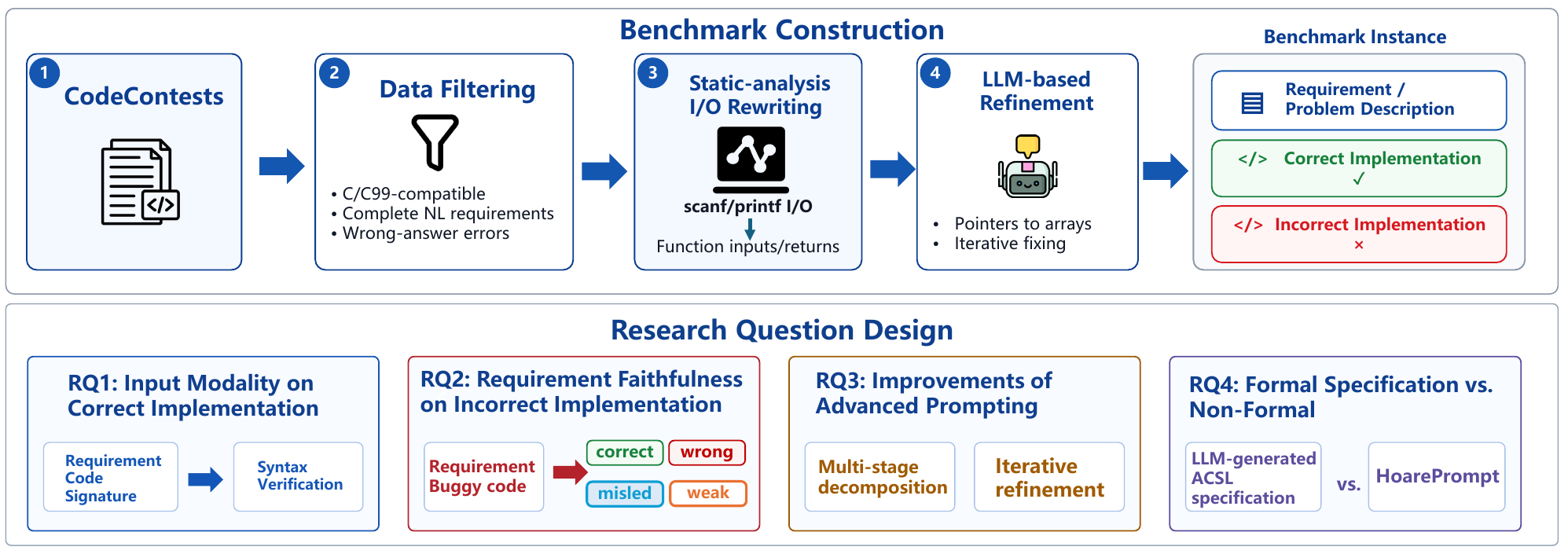}
\caption{Overview of our study methodology.}
\label{fig:methodology-overview}
\end{figure*}

\subsection{Overview and Task Definition}

The overview of our study is shown in Figure~\ref{fig:methodology-overview}. In this study, we focus on the C programming language and ACSL specification language for formal verification.
Given a natural language requirement $R$ and an implementation $I$ whose correctness is unknown, the bimodal specification generation task asks an LLM to generate a formal specification $S$:
\begin{equation}
    G(R, I) \rightarrow S,
    \quad
    S = \langle P, Q, A \rangle .
\end{equation}
Here, $P$ denotes the pre-condition, $Q$ denotes the post-condition, and $A$ denotes auxiliary specification elements such as loop invariants and assertions. Unlike code-only specification generation, the semantic oracle of this task is the requirement, not the implementation. The implementation should provide proof-relevant context, such as variable names, control flow, and loop structure, but the generated specification should describe the expected behavior in $R$.

We denote the behaviors allowed by the requirement, implementation, and specification as $\mathsf{Beh}(R)$, $\mathsf{Beh}(I)$, and $\mathsf{Beh}(S)$, respectively. A requirement-faithful specification should satisfy $\mathsf{Beh}(S)=\mathsf{Beh}(R)$. The verifier then checks whether $I$ satisfies $S$, written as $\mathsf{Verify}(I,S)$. For a correct implementation, a faithful specification should pass verification. For an incorrect implementation, a faithful specification should fail verification by exposing the mismatch between $I$ and $R$.

Therefore, \textbf{verification pass is not always a success signal} in the bimodal setting. If an incorrect implementation passes verification, the generated specification may be too weak, or it may be misled by the implementation and encode the wrong behavior. This formalization motivates our evaluation on both correct and incorrect programs.

\subsection{Benchmark Construction}

We create our evaluation benchmark from the CodeContests dataset~\cite{li2022competition}. CodeContests consists of 10,000 programming contest problems from multiple online programming platforms. Each problem is associated with a problem description, test cases, and multiple solutions in C++, Python, and Java. The solutions contain both correct and incorrect implementations written by \textbf{human programmers}. We perform the following steps on CodeContests to curate our benchmark:

\begin{enumerate}
    \item \textbf{Data filtering:} We filter the problems with C++ solutions that are compatible with the C99 standard, to support ACSL and the Frama-C verifier. 
    We also filter the natural language problem descriptions that are incomplete, e.g., the key requirements are described in pictures or external links.
    For incorrect implementations, we only choose those that belong to the ``wrong answer'' category, because these programs match our goal of detecting behavior deviation from the requirements using formal specifications, while removing programs that have timeout or other runtime errors.

    \item \textbf{Static analysis-based I/O rewriting:} The original I/O of CodeContests C++ solutions is from \texttt{scanf} and \texttt{printf}, which is not suitable for verification. In ACSL and Frama-C, the pre-conditions and post-conditions are used to describe the function input arguments and return values, so we first remove programs with non-uniform I/O formats (e.g., the program may output different types of values in different cases), and then rewrite the I/O into function input arguments and return values. 
    Since most solutions in CodeContests do not have a \texttt{main} function with input arguments, we create a new function \texttt{\_\_solve} to pack the core logic of the original implementation, the input arguments and return values are assigned to the \texttt{\_\_solve} function.
    
    \item \textbf{LLM-based refinement:} For all solutions to the same problem, we perform LLM-based rewriting to unify their I/O formats. We also convert pointers to arrays with variable lengths by adding new input arguments representing array lengths, e.g., translate \texttt{foo(int *a)} to \texttt{foo(int n, int a[n])}.
    After the rewriting, we run these programs with the provided test cases to make sure that the new return values are identical to the original program. If the rewritten program triggers runtime errors, we further perform a maximum of 3 rounds of iterative fixing by providing the error report to the LLM. We use the gpt-5.4-mini LLM for this rewriting step.
\end{enumerate}

After the filtering, there remain 1,083 programming problems. The rewriting steps make the programs in our benchmark structurally different from the original CodeContests, which further reduces the risk of data leakage.
From the 1,083 problems, we further select the first 500 problems for evaluation efficiency. In each problem, we select at most 1 correct implementation, and 2 incorrect implementations to form our final benchmark of 1403 programs. This is so far the first benchmark for formal specification generation that covers both natural language and program language sources, and allows verification on both correct and incorrect programs. The detailed statistics of our benchmark are shown in Table~\ref{tab:benchmark-statistics}.

\begin{table*}[t]
\centering
\caption{Statistics of our benchmark. CF-x denotes different complexity levels measured by control flow structures. CF-1: programs without branches, CF-2: programs with branches but no loops, CF-3: programs with single-level loops, and CF-4: programs with nested loops.}
\label{tab:benchmark-statistics}
\scriptsize
\setlength{\tabcolsep}{3pt}
\begin{tabular}{lrrrr|rrrr|rr}
\hline
\multirow{2}{*}{Category} &
\multicolumn{4}{c|}{Answer Statistics} &
\multicolumn{4}{c|}{Control-flow Complexity} &
\multicolumn{2}{c}{Function Definition Type} \\
\cline{2-5}\cline{6-9}\cline{10-11}
& \#Problems & \#Programs & Avg. LOC & Avg. CC
& CF-1 & CF-2 & CF-3 & CF-4
& Main-only & Custom func. \\
\hline
Overall & 500 & 1403 & 32.8 & 6.23 & 120 (8.5\%) & 180 (12.8\%) & 624 (44.5\%) & 479 (34.2\%) & 1123 (80.0\%) & 280 (20.0\%) \\
Correct & -- & 500 & 31.5 & 6.04 & 48 (9.6\%) & 59 (11.8\%) & 221 (44.2\%) & 172 (34.4\%) & 387 (77.4\%) & 113 (22.6\%) \\
Incorrect & -- & 903 & 33.4 & 6.33 & 72 (8.0\%) & 121 (13.4\%) & 403 (44.6\%) & 307 (34.0\%) & 736 (81.5\%) & 167 (18.5\%) \\
\hline
\end{tabular}
\end{table*}

\subsection{Research Questions}

\subsubsection{RQ1: input modality}

In RQ1, we compare three input forms on correct implementations: requirement + complete code, code only, and requirement + function signature, where the signature only constrains the generated pre-/post-condition format. We report syntactic correctness and verification pass rates.

\subsubsection{RQ2: performance on incorrect implementations}

In RQ2, we provide the natural language requirement and an incorrect implementation, without telling the LLM that the code is incorrect, and ask it to generate an ACSL specification for checking the implementation against the requirement. We report syntactic correctness and verification \textbf{failure} rates, because a requirement-consistent specification should fail on an incorrect implementation.
Since verification failure may also come from an inconsistent or overly strong specification, and verification pass may indicate either a misled or weak specification, we further classify generated specifications as follows: \emph{Correct} accurately describes the requirement and fails verification; \emph{Wrong} fails verification but is inconsistent with the requirement; \emph{Misled} passes verification by following the incorrect implementation; \emph{Weak} passes because it is too weak to expose the bug; and \emph{Other} covers cases that cannot be reliably classified.

Because direct semantic classification is hard to automate, we approximate specification behavior using test case-based runtime checking with Frama-C E-ACSL. For each specification, we run up to 40 CodeContests test cases, prioritizing cases on which the original incorrect implementation fails, and compare the original execution result with the E-ACSL-instrumented execution:

\begin{itemize}
    \item \textbf{A/B:} the original implementation passes the test, and E-ACSL either terminates normally (A) or aborts due to a specification violation (B).

    \item \textbf{C/D:} the original implementation fails the test, and E-ACSL either terminates normally (C) or aborts due to a specification violation (D).

    \item \textbf{E:} E-ACSL execution exceeds the runtime limit.
\end{itemize}

We compute the following rates for classification:
\begin{IEEEeqnarray}{lCl}
	\mathsf{ValidRate}       & = & \frac{A+B+C+D+E}{N} \\[2pt]
	\mathsf{WrongAcceptRate} & = & \frac{C}{C+D} \\[2pt]
	\mathsf{RightAbortRate}  & = & \frac{B}{A+B} \\[2pt]
	\mathsf{TimeoutRate}     & = & \frac{E}{A+B+C+D+E}
\end{IEEEeqnarray}
where $N$ is the number of selected test cases; tests that cannot be executed by E-ACSL are counted in $N$ but not in A-E. $\mathsf{WrongAcceptRate}$ measures accepted failing executions, $\mathsf{RightAbortRate}$ measures rejected passing executions, and unavailable denominators are classified as \emph{Other}.

We combine these rates with the Frama-C verification result to classify generated specifications for incorrect implementations, as summarized in Table~\ref{tab:eacsl-classification}.

\begin{table}[t]
\centering
\caption{E-ACSL-based classification criteria for incorrect implementations.}
\label{tab:eacsl-classification}
\scriptsize
\setlength{\tabcolsep}{5pt}
\resizebox{\columnwidth}{!}{%
\begin{tabular}{l c c c c @{\hspace{4pt}}c@{\hspace{4pt}} c}
\hline
Category & Verification Result & ValidRate & TimeoutRate & WrongAcceptRate & & RightAbortRate \\ 
\hline
Correct & Fail & $\ge \theta_v$ & $< \theta_t$ & $< \theta_w$ & --- & --- \\
Wrong   & Fail & $\ge \theta_v$ & $< \theta_t$ & $\ge \theta_w$ & --- & --- \\
Weak    & Pass & $\ge \theta_v$ & $< \theta_t$ & $\ge \theta_w$ & and & $\le \theta_r$ \\
Misled  & Pass & $\ge \theta_v$ & $< \theta_t$ & $< \theta_w$ & or  & $> \theta_r$ \\
Other   & Fail / Pass & $< \theta_v$ & $\ge \theta_t$ & --- & --- & --- \\
\hline
\end{tabular}%
}
\end{table}

Here, $\theta_v$, $\theta_t$, $\theta_w$, and $\theta_r$ are the calibrated thresholds for $\mathsf{ValidRate}$, $\mathsf{TimeoutRate}$, $\mathsf{WrongAcceptRate}$, and $\mathsf{RightAbortRate}$; their values are calibrated by the human study in Table~\ref{tab:rq2-eacsl-human-consistency}.

\subsubsection{RQ3: advanced prompting}

Previous RQs use basic one-pass prompting. In RQ3, we evaluate whether task decomposition and verifier-feedback refinement improve bimodal specification generation, especially when the implementation is inconsistent with the requirement.

To study this question, we propose \textsc{RULER} (\textbf{R}equirement \textbf{U}nderstanding with \textbf{L}LM-judg\textbf{e}d \textbf{R}efinement), a two-stage workflow. \textsc{RULER} first asks the LLM to reason about expected pre-/post-conditions from the natural-language requirement alone, then uses the requirement, implementation, verifier feedback, and an LLM judge to generate and iteratively refine the ACSL specification.
In each refinement round, the generator produces a specification, Frama-C returns a verification result or error message, and the judge decides whether the current specification should be further refined. Unlike prior verifier-feedback methods~\cite{ma2025specgen,le2025can} that stop when verification passes, \textsc{RULER} can stop even when verification fails if the judge considers the specification requirement-consistent, because verification pass is not a reliable success signal on incorrect implementations.
We also evaluate AutoSpec~\cite{wen2024enchanting} without its refinement step, which optimizes specifications toward successful verification. This adaptation allows us to evaluate specification generation on both correct and incorrect implementations.

\subsubsection{RQ4: comparison with non-formal approaches}
Recently, there has been an interest in studying the misalignment between natural language requirements and actual program behavior~\cite{konstantinou2024llms,bouras2025hoareprompt,tian2025aligning,florian2026inferring} in LLM-driven software engineering tasks. 
Among the works that focus on predicting program correctness from natural language requirements~\cite{bouras2025hoareprompt,florian2026inferring}, HoarePrompt~\cite{bouras2025hoareprompt} is the closest to our setting, because it is also \textbf{inspired by formal methods}, but performs the reasoning in natural language rather than machine-checkable formal specifications. 
HoarePrompt propagates natural-language state descriptions along the control-flow graph, applies few-shot-driven loop unrolling, annotates inferred states into code, and uses an LLM classifier to predict whether implementation $I$ satisfies requirement $R$.

In RQ4, we reimplement HoarePrompt for C programs and compare it with our specification-generation pipelines (basic prompting and \textsc{RULER}) as program-level correctness classifiers. For our pipelines, Frama-C verification success is treated as \emph{Correct}, and all other outcomes as \emph{Incorrect}. We follow~\cite{bouras2025hoareprompt} and report the balanced accuracy (BA), true positive rate (TPR), false positive rate (FPR), and MCC on our benchmark.

\section{Evaluation}
\subsection{Evaluation Setup}

We evaluate three LLMs for ACSL specification generation: GPT-5.5, GPT-5.4-mini, and DeepSeek-v4-pro. These models cover both flagship proprietary models and more cost-efficient models, allowing us to compare whether model capability affects the faithfulness and verifiability of generated specifications. 
We disable the thinking mode of these LLMs to generate concise answers.

All generated ACSL specifications are verified with Frama-C 32.0. Before running verification, we first validate syntactic correctness using the Frama-C kernel parser and type checker, invoked with \texttt{frama-c -typecheck}. 
We use Frama-C's WP plug-in (\texttt{-wp}) to generate weakest-precondition proof obligations. The proof obligations are discharged by the SMT solvers Alt-Ergo and Z3. We set the verification timeout to 30 seconds. 
A program-specification pair is considered verified only when all proof obligations are discharged within the timeout. Verification errors and timeout cases are both counted as verification failures.

\subsection{RQ1: input modality}

Table~\ref{tab:rq1-input-modality} summarizes the results for RQ1, where LLMs generate specifications from correct code implementations with different input modalities. The evaluation is performed on 500 correct code implementations with three input settings: requirement with code (\emph{Req.+Code}), requirement with function signature only (\emph{Req.+Sig.}), and code only.

\begin{table*}[t]
\centering
\caption{Syntax checking and verification results under different input modalities on correct implementations.}
\label{tab:rq1-input-modality}
\scriptsize
\setlength{\tabcolsep}{2pt}
\renewcommand{\arraystretch}{1.05}
\resizebox{\textwidth}{!}{
\begin{tabular}{ll|ccc|ccc|ccc}
\hline
\multirow{2}{*}{Scope} & \multirow{2}{*}{Model} &
\multicolumn{3}{c|}{Req.+Code} &
\multicolumn{3}{c|}{Req.+Sig.} &
\multicolumn{3}{c}{Code only} \\
\cline{3-11}
& & Syn. & Ver. & Ver.|Syn. & Syn. & Ver. & Ver.|Syn. & Syn. & Ver. & Ver.|Syn. \\
\hline
\multirow{3}{*}{Overall}
& gpt-5.4-mini & 269 (53.80\%) & 73 (14.60\%) & 73 (27.14\%) & 359 (71.80\%) & 41 (8.20\%) & 41 (11.42\%) & 243 (48.60\%) & 74 (14.80\%) & 74 (30.45\%) \\
& gpt-5.5 & 384 (76.80\%) & 126 (25.20\%) & 126 (32.81\%) & 185 (37.00\%) & 52 (10.40\%) & 52 (28.11\%) & 400 (80.00\%) & 183 (36.60\%) & 183 (45.75\%) \\
& deepseek-v4-pro & 253 (50.60\%) & 78 (15.60\%) & 78 (30.83\%) & 165 (33.00\%) & 46 (9.20\%) & 46 (27.88\%) & 250 (50.00\%) & 84 (16.80\%) & 84 (33.60\%) \\
\hline
\multirow{3}{*}{CF-1}
& gpt-5.4-mini & 41 (85.42\%) & 29 (60.42\%) & 29 (70.73\%) & 45 (93.75\%) & 23 (47.92\%) & 23 (51.11\%) & 39 (81.25\%) & 28 (58.33\%) & 28 (71.79\%) \\
& gpt-5.5 & 46 (95.83\%) & 31 (64.58\%) & 31 (67.39\%) & 43 (89.58\%) & 29 (60.42\%) & 29 (67.44\%) & 44 (91.67\%) & 33 (68.75\%) & 33 (75.00\%) \\
& deepseek-v4-pro & 41 (85.42\%) & 27 (56.25\%) & 27 (65.85\%) & 40 (83.33\%) & 26 (54.17\%) & 26 (65.00\%) & 40 (83.33\%) & 27 (56.25\%) & 27 (67.50\%) \\
\hline
\multirow{3}{*}{CF-2}
& gpt-5.4-mini & 51 (86.44\%) & 30 (50.85\%) & 30 (58.82\%) & 52 (88.14\%) & 16 (27.12\%) & 16 (30.77\%) & 44 (74.58\%) & 31 (52.54\%) & 31 (70.45\%) \\
& gpt-5.5 & 59 (100.00\%) & 35 (59.32\%) & 35 (59.32\%) & 46 (77.97\%) & 22 (37.29\%) & 22 (47.83\%) & 58 (98.31\%) & 47 (79.66\%) & 47 (81.03\%) \\
& deepseek-v4-pro & 50 (84.75\%) & 31 (52.54\%) & 31 (62.00\%) & 44 (74.58\%) & 18 (30.51\%) & 18 (40.91\%) & 50 (84.75\%) & 31 (52.54\%) & 31 (62.00\%) \\
\hline
\multirow{3}{*}{CF-3}
& gpt-5.4-mini & 114 (51.58\%) & 14 (6.33\%) & 14 (12.28\%) & 148 (66.97\%) & 2 (0.90\%) & 2 (1.35\%) & 89 (40.27\%) & 14 (6.33\%) & 14 (15.73\%) \\
& gpt-5.5 & 167 (75.57\%) & 55 (24.89\%) & 55 (32.93\%) & 56 (25.34\%) & 1 (0.45\%) & 1 (1.79\%) & 179 (81.00\%) & 79 (35.75\%) & 79 (44.13\%) \\
& deepseek-v4-pro & 107 (48.42\%) & 19 (8.60\%) & 19 (17.76\%) & 55 (24.89\%) & 2 (0.90\%) & 2 (3.64\%) & 112 (50.68\%) & 24 (10.86\%) & 24 (21.43\%) \\
\hline
\multirow{3}{*}{CF-4}
& gpt-5.4-mini & 63 (36.63\%) & 0 (0.00\%) & 0 (0.00\%) & 114 (66.28\%) & 0 (0.00\%) & 0 (0.00\%) & 71 (41.28\%) & 1 (0.58\%) & 1 (1.41\%) \\
& gpt-5.5 & 112 (65.12\%) & 5 (2.91\%) & 5 (4.46\%) & 40 (23.26\%) & 0 (0.00\%) & 0 (0.00\%) & 119 (69.19\%) & 24 (13.95\%) & 24 (20.17\%) \\
& deepseek-v4-pro & 55 (31.98\%) & 1 (0.58\%) & 1 (1.82\%) & 26 (15.12\%) & 0 (0.00\%) & 0 (0.00\%) & 48 (27.91\%) & 2 (1.16\%) & 2 (4.17\%) \\
\hline
\end{tabular}
}
\end{table*}

The results show that the input modality has a substantial impact on both syntactic correctness and verification success. Aggregating the three models, \emph{Code only} achieves the highest overall verification rate, verifying 341 out of 1,500 generated specifications (22.73\%). This is higher than \emph{Req.+Code}, which verifies 277 specifications (18.47\%), and more than twice the rate of \emph{Req.+Sig.}, which verifies only 139 specifications (9.27\%). The same pattern holds when considering only syntactically correct specifications: \emph{Code only} reaches 38.19\% Ver.|Syn., compared with 30.57\% for \emph{Req.+Code} and 19.61\% for \emph{Req.+Sig.}.

\begin{figure}[t]
\centering
\includegraphics[width=\columnwidth]{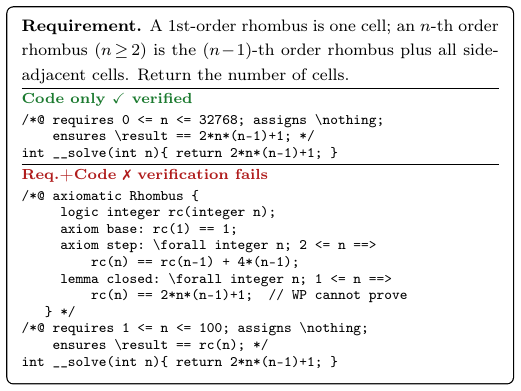}
\caption{A correct implementation where \emph{Code only}
verifies but \emph{Req.+Code} does not (GPT-5.5). The requirement defines
the rhombus \emph{recursively}, leading the model to an inductive
\texttt{axiomatic} specification whose closed-form \texttt{lemma} WP cannot
discharge; the code-only specification mirrors the implementation and is proved immediately.}
\label{fig:rq1-case}
\end{figure}

A first observation is that providing both the requirement and the implementation does not consistently improve verifiability over providing the implementation alone. Since RQ1 uses correct implementations, the code already encodes behavior consistent with the requirement and provides concrete operational structure that helps LLMs generate specifications Frama-C can prove.
Figure~\ref{fig:rq1-case} illustrates this effect. Under \emph{Req.+Code}, the recursive requirement leads GPT-5.5 to encode the rhombus definition as an \texttt{axiomatic} block and a logic-function postcondition. This specification is not necessarily wrong, and may be more faithful to the requirement, but proving it requires an induction-style \texttt{lemma} beyond what WP and the back-end SMT solvers discharge automatically. Under \emph{Code only}, the model instead mirrors the implementation's closed-form postcondition, which WP proves immediately. Thus, the requirement can steer the model toward requirement-fitting but proof-hostile specifications.

A second observation is that \emph{Req.+Sig.} is consistently the weakest modality. Although the function signature constrains the interface of the generated specification, it hides the implementation details needed to infer precise memory accesses, boundary cases, loop behavior, and intermediate invariants. This gap is especially visible for complex programs: under CF-3, \emph{Req.+Sig.} verifies only 5 out of 663 specifications (0.75\%), and under CF-4 it verifies none. Thus, natural-language requirements plus interface information are insufficient for reliable ACSL generation in our benchmark.

The results also reveal a large gap between syntactic correctness and verification success: across all modalities, the majority of syntactically valid specifications still fail WP discharge (Table~\ref{tab:rq1-input-modality}). Syntactic validity is therefore a necessary but weak indicator of specification quality.

Control-flow complexity further amplifies these differences. For CF-1 and CF-2, \emph{Code only} verifies 61.11\% and 61.58\% of generated specifications, respectively. The rate drops to 17.65\% for CF-3 and 5.23\% for CF-4. The same downward trend appears for the other modalities. This suggests that LLMs can often generate usable contracts for simpler programs, but struggle to produce sufficiently strong and proof-friendly specifications when more complex control flow is involved.

At the model level, GPT-5.5 obtains the best overall performance, especially under \emph{Code only}, where it verifies 183 out of 500 specifications (36.60\%). However, model capability does not eliminate the modality effect: GPT-5.5 still drops to 10.40\% verification under \emph{Req.+Sig.}. Therefore, stronger models improve absolute performance, but the availability of implementation-level information remains a decisive factor.


\noindent\textbf{Answer to RQ1.}
Input modality strongly affects LLM-generated ACSL specifications. On correct implementations, code-only prompting achieves the highest verification rate, while requirement-plus-signature prompting performs the worst. This suggests that implementation structure is currently more useful than natural-language requirements for producing verifiable ACSL specifications.

\subsection{RQ2: performance on incorrect implementations}

Before applying the E-ACSL-based taxonomy to all incorrect implementations, we first examine whether the calibrated thresholds produce labels that are consistent with human analysis.

\begin{table}[t]
\centering
\caption{Consistency between test case-based E-ACSL classification and human analysis across three models (50 specifications per model).}
\label{tab:rq2-eacsl-human-consistency}
\scriptsize
\setlength{\tabcolsep}{2pt}
\renewcommand{\arraystretch}{1.05}
\begin{tabular}{llrrrrr}
\toprule
Model & Analysis & Correct & Wrong & Misled & Weak & Other \\
\midrule
gpt-5.4-mini & Human & 14 & 9 & 14 & 11 & 2 \\
& E-ACSL & 12 & 4 & 12 & 13 & 9 \\
\midrule
gpt-5.5 & Human & 15 & 8 & 19 & 6 & 2 \\
& E-ACSL & 7 & 5 & 19 & 6 & 13 \\
\midrule
deepseek-v4-pro & Human & 12 & 10 & 15 & 10 & 3 \\
& E-ACSL & 10 & 7 & 14 & 11 & 8 \\
\bottomrule
\end{tabular}
\end{table}

We compare E-ACSL classifications with human analysis on 50 specifications from each of the three models. Table~\ref{tab:rq2-eacsl-human-consistency} shows similar agreement patterns across models. Cohen's kappa is 0.5425, 0.5083, and 0.5677 for GPT-5.4-mini, GPT-5.5, and DeepSeek-v4-pro, respectively; excluding \emph{Other} cases, it increases to 0.6751, 0.6718, and 0.7166. E-ACSL assigns more cases to \emph{Other} than human analysis for all three models, indicating that the automated pipeline leaves more specifications unclassified.

Based on this calibration, we set $\theta_v=0.5$, $\theta_t=0.6$, $\theta_w=0.6$, and $\theta_r=0.10$ for the full RQ2 analysis. These thresholds are then used to produce the E-ACSL-based taxonomy reported in Table~\ref{tab:rq2-incorrect}.

The classifications are stable under the tested threshold perturbations. We vary each threshold individually while keeping the others fixed. Table~\ref{tab:rq2-threshold-sensitivity} reports the maximum label-change rate for each threshold over the two tested values. Across all settings, at most 2.06\% of classifications change, with \textsf{WrongAcceptRate} producing the largest changes.

\begin{table}[t]
\centering
\caption{Maximum label-change rates under individual threshold perturbations, measured over syntactically valid specifications.}
\label{tab:rq2-threshold-sensitivity}
\scriptsize
\setlength{\tabcolsep}{2pt}
\renewcommand{\arraystretch}{1.05}
\begin{tabular}{lcrrr}
\toprule
Threshold & Tested values & gpt-5.4-mini & gpt-5.5 & deepseek-v4-pro \\
\midrule
$\theta_v$ & 0.4, 0.6 & 0.00\% & 0.00\% & 0.00\% \\
$\theta_t$ & 0.5, 0.7 & 0.00\% & 0.00\% & 0.19\% \\
$\theta_w$ & 0.5, 0.7 & 2.04\% & 2.06\% & 1.75\% \\
$\theta_r$ & 0.0, 0.2 & 1.22\% & 0.28\% & 1.17\% \\
\bottomrule
\end{tabular}
\end{table}

\begin{table*}[t]
\centering
\caption{Specification verification results and detailed taxonomy on incorrect implementations under requirement-with-code input.}
\label{tab:rq2-incorrect}
\scriptsize
\setlength{\tabcolsep}{3pt}
\renewcommand{\arraystretch}{1.05}
\resizebox{0.9\textwidth}{!}{
\begin{tabular}{llccc@{\hspace{4pt}}c@{\hspace{4pt}}crrrrc}
\toprule
\multirow{2}{*}{Scope} & \multirow{2}{*}{Model} & \multicolumn{3}{c}{Verification Outcome} & & \multicolumn{5}{c}{Detailed Taxonomy} & \multirow{2}{*}{\begin{tabular}{c}Final Correct\\Rate\end{tabular}} \\
\cmidrule(lr){3-5}\cmidrule(lr){7-11}
& & Syn. & Ver. & Ver.|Syn. & & Correct/Syn. & Wrong & Misled & Weak & Other & \\
\midrule
\multirow{3}{*}{Overall}
& gpt-5.4-mini & 490 (54.26\%) & 407 (45.07\%) & 407 (83.06\%) & & 222 (45.31\%) & 103 & 48 & 35 & 82 & 24.58\% \\
& gpt-5.5 & 727 (80.51\%) & 694 (76.85\%) & 694 (95.46\%) & & 235 (32.32\%) & 156 & 27 & 6 & 303 & 26.02\% \\
& deepseek-v4-pro & 513 (56.81\%) & 444 (49.17\%) & 444 (86.55\%) & & 219 (42.69\%) & 109 & 36 & 33 & 116 & 24.25\% \\
\midrule
\multirow{3}{*}{CF-1}
& gpt-5.4-mini & 67 (93.06\%) & 38 (52.78\%) & 38 (56.72\%) & & 18 (26.87\%) & 3 & 17 & 12 & 17 & 25.00\% \\
& gpt-5.5 & 71 (98.61\%) & 62 (86.11\%) & 62 (87.32\%) & & 27 (38.03\%) & 4 & 7 & 2 & 31 & 37.50\% \\
& deepseek-v4-pro & 72 (100.00\%) & 50 (69.44\%) & 50 (69.44\%) & & 21 (29.17\%) & 3 & 11 & 11 & 26 & 29.17\% \\
\midrule
\multirow{3}{*}{CF-2}
& gpt-5.4-mini & 107 (88.43\%) & 73 (60.33\%) & 73 (68.22\%) & & 39 (36.45\%) & 17 & 21 & 13 & 17 & 32.23\% \\
& gpt-5.5 & 113 (93.39\%) & 104 (85.95\%) & 104 (92.04\%) & & 47 (41.59\%) & 16 & 9 & 0 & 41 & 38.84\% \\
& deepseek-v4-pro & 105 (86.78\%) & 81 (66.94\%) & 81 (77.14\%) & & 44 (41.90\%) & 15 & 11 & 13 & 22 & 36.36\% \\
\midrule
\multirow{3}{*}{CF-3}
& gpt-5.4-mini & 202 (50.12\%) & 186 (46.15\%) & 186 (92.08\%) & & 96 (47.52\%) & 59 & 8 & 8 & 31 & 23.82\% \\
& gpt-5.5 & 336 (83.37\%) & 321 (79.65\%) & 321 (95.54\%) & & 90 (26.79\%) & 89 & 11 & 4 & 142 & 22.33\% \\
& deepseek-v4-pro & 233 (57.82\%) & 211 (52.36\%) & 211 (90.56\%) & & 102 (43.78\%) & 62 & 14 & 8 & 47 & 25.31\% \\
\midrule
\multirow{3}{*}{CF-4}
& gpt-5.4-mini & 114 (37.13\%) & 110 (35.83\%) & 110 (96.49\%) & & 69 (60.53\%) & 24 & 2 & 2 & 17 & 22.48\% \\
& gpt-5.5 & 207 (67.43\%) & 207 (67.43\%) & 207 (100.00\%) & & 71 (34.30\%) & 47 & 0 & 0 & 89 & 23.13\% \\
& deepseek-v4-pro & 103 (33.55\%) & 102 (33.22\%) & 102 (99.03\%) & & 52 (50.49\%) & 29 & 0 & 1 & 21 & 16.94\% \\
\bottomrule
\end{tabular}
}
\end{table*}

Table~\ref{tab:rq2-incorrect} shows that high verification-failure rates do not imply high requirement faithfulness. Since all implementations in RQ2 are incorrect, a requirement-consistent specification should expose the implementation error and fail verification. However, among syntactically correct specifications, only about one third to one half are classified as correct, and the final correct rate over all incorrect implementations is even lower, around one quarter for every model. This result indicates that verifier failure is only a coarse signal. Many specifications fail verification for reasons unrelated to faithfully capturing the requirement, such as being semantically wrong, too complex to classify, or inconclusive under the checker.

A second observation is that model capability improves syntactic validity and verification-failure rates, but does not proportionally improve requirement faithfulness. GPT-5.5 generates substantially more syntactically correct specifications and achieves the highest verification-failure rate among them, yet only about a third of these are classified as correct, and its final correct rate is only slightly higher than the other two models. This gap suggests that stronger models are better at producing ACSL accepted by Frama-C and specifications that trigger proof failures, but these improvements do not directly translate into requirement-consistent specifications.

False positives remain a smaller but important failure mode. A false positive occurs when a generated specification passes verification on an incorrect implementation. These cases are classified as either \emph{Misled} or \emph{Weak}. Although far fewer than verification-failure cases, they are especially relevant to the goal of formal verification: the verifier would accept an incorrect implementation under the generated specification. Moreover, misled specifications dominate weak ones across all three models, accounting for over half of all false positives (and 81.82\% for GPT-5.5). This pattern suggests that the main risk is not only that LLMs write underspecified contracts, but that they may actively align the specification with the erroneous behavior of the implementation.

The large number of \emph{Other} cases for GPT-5.5 (41.68\% of its syntactically correct specifications) requires cautious interpretation. GPT-5.5 tends to generate more complex specifications with auxiliary predicates, logic functions, or axioms that may exceed our E-ACSL-based checking and manual classification, so some \emph{Other} cases may be valid specifications our pipeline cannot fully validate. We nevertheless conservatively exclude them from the final correct rate, as they do not provide reliable evidence of requirement faithfulness on the incorrect implementation.

Control-flow complexity further changes the nature of the failures. For CF-1 and CF-2 programs, syntactic correctness is high across models, but false positives remain visible, especially for GPT-5.4-mini and DeepSeek-v4-pro. For CF-3 and CF-4 programs, verification failures become more common, but the fraction of correct specifications does not increase accordingly. For example, GPT-5.5 reaches 95.54\% Ver.|Syn. on CF-3 and 100.00\% on CF-4, yet its Correct/Syn. rates are only 26.79\% and 34.30\%, respectively. This shows that more complex control flow makes specifications harder to validate and interpret: failure of verification becomes easier to obtain, but harder to treat as evidence of requirement faithfulness.

At the same time, implementation-induced misalignment is not limited to complex programs. In CF-1 and CF-2, where syntactic correctness is already high, false positives still appear across all models. Many of these cases are misled rather than merely weak; for example, GPT-5.4-mini produces 17 misled specifications in CF-1 and 21 in CF-2, compared with 12 and 13 weak specifications. This suggests that even for structurally simple programs, LLMs may align the specification with the incorrect implementation rather than the natural-language requirement.

\noindent\textbf{Answer to RQ2.}
LLMs can generate requirement-consistent specifications for some incorrect implementations, but the success rate remains low. Although many generated specifications fail verification on incorrect code, only about one quarter of all incorrect implementations receive a specification that is both syntactically correct and classified as requirement-consistent. False positives are less frequent, but they reveal a critical risk: LLM-generated specifications are more often misled by the incorrect implementation than merely too weak. Stronger models improve syntactic validity and verification-failure rates, but they also produce more complex inconclusive specifications, suggesting that future work should study how to prevent implementation-misled specifications and how to validate complex LLM-generated ACSL specifications more reliably.

\subsection{RQ3: advanced prompting}

Table~\ref{tab:rq3-advanced-overall} compares the basic prompt, AutoSpec, \textsc{RULER} (2-stage), and \textsc{RULER} (2-stage+refine). We use the final \emph{Success} rate as the primary metric, because it directly measures whether a generated specification is syntactically accepted and successfully validated under the corresponding implementation setting.

AutoSpec does not improve the final success rate over basic prompting. Its success rates are 13.80\%--17.00\% on correct implementations and 18.94\%--24.92\% on incorrect implementations, with lower rates for all models except a tie for DeepSeek-v4-pro on correct implementations.

The \textsc{RULER} (2-stage) strategy alone does not consistently improve specification generation. On correct implementations, the overall number of successful specifications decreases from 277 under the basic prompt to 248 under \textsc{RULER} (2-stage), reducing the final success rate from 18.47\% to 16.53\%. This decrease appears for all three models: GPT-5.4-mini drops from 73 to 66 successful specifications, GPT-5.5 from 126 to 110, and DeepSeek-v4-pro from 78 to 72. On incorrect implementations, \textsc{RULER} (2-stage) yields only a marginal aggregate change, from 676 to 680 successful specifications. This small overall gain is mainly due to GPT-5.5, whose successful cases increase from 235 to 256, while GPT-5.4-mini and DeepSeek-v4-pro slightly decrease. These results suggest that decomposing the task into two stages is not sufficient by itself. Although task decomposition can make the prompting process more structured, it does not reliably help the model produce specifications that pass the final validation.

Adding iterative refinement substantially improves the final success rate on correct implementations. Compared with the basic prompt, \textsc{RULER} (2-stage+refine) increases the total number of successful specifications from 277 to 364, improving the final success rate from 18.47\% to 24.27\%. The improvement is consistent across all three models. GPT-5.4-mini improves from 73 to 88 successful specifications, GPT-5.5 from 126 to 166, and DeepSeek-v4-pro from 78 to 110. Compared with \textsc{RULER} (2-stage), refinement also recovers and exceeds the performance loss introduced by task decomposition. This result indicates that verifier-guided feedback is useful when the implementation is correct: failed verification provides a relatively direct signal that the generated specification may need to be repaired, strengthened, or rewritten.

The effect of iterative refinement is less stable on incorrect implementations. In aggregate, \textsc{RULER} (2-stage+refine) improves over the basic prompt, increasing the number of successful specifications from 676 to 799 and the final success rate from 24.95\% to 29.49\%. It also improves over \textsc{RULER} (2-stage) at the aggregate level, from 680 to 799 successful cases. However, this aggregate improvement hides model-level instability. GPT-5.4-mini improves from 220 successful cases under \textsc{RULER} (2-stage) to 291 after refinement, and DeepSeek-v4-pro improves from 204 to 257. In contrast, GPT-5.5 decreases from 256 successful cases under \textsc{RULER} (2-stage) to 251 after refinement. Thus, while refinement generally helps on incorrect implementations, it does not uniformly improve every model.

This difference between correct and incorrect implementations reveals an important limitation of refinement-based prompting. For correct implementations, the target behavior and the implementation behavior are aligned, so verification feedback is a useful signal for repairing the specification. For incorrect implementations, however, the model must distinguish the natural-language requirement from the behavior of the faulty code. Refinement can therefore become ambiguous: a verification result may encourage the model to adjust the specification toward the implementation, even when the implementation is wrong. The GPT-5.5 result on incorrect implementations illustrates this risk. Although GPT-5.5 benefits substantially from refinement on correct implementations, its success count decreases after refinement on incorrect implementations compared with \textsc{RULER} (2-stage). This suggests that stronger models and iterative feedback do not automatically solve the requirement-alignment problem identified in RQ2.

Overall, RQ3 shows that the full advanced prompting strategy is useful, but its benefit comes mainly from combining task decomposition with iterative repair rather than from decomposition alone. \textsc{RULER} (2-stage+refine) improves the final success rate on both correct and incorrect implementations compared with the basic prompt. At the same time, the weaker and less stable gains on incorrect implementations indicate that LLMs still struggle to judge whether a specification is consistent with the requirement or merely consistent with the given implementation.

\begin{figure*}[t]
\centering
\includegraphics[width=\textwidth]{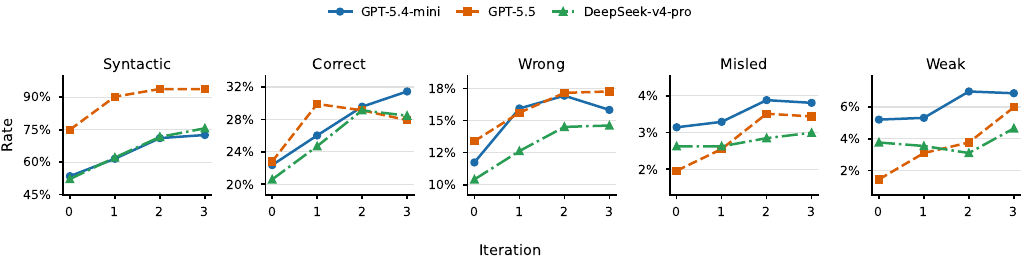}
\caption{Evolution of specification categories during iterative repair on incorrect implementations. The x-axis value 0 denotes the output before iterative repair, and 1--3 denote the outputs after each repair iteration. Each subplot uses an independent y-axis scale and reports rates over all incorrect implementations.}
\label{fig:rq3-iter-incorrect}
\end{figure*}

Figure~\ref{fig:rq3-iter-incorrect} further shows how iterative repair changes generated specifications over repair rounds on incorrect implementations. Syntactic correctness increases, and the rate of requirement-consistent specifications improves for some models. However, wrong, misled, and weak specifications do not consistently decrease. In particular, GPT-5.5 reaches its highest correct rate after the first repair iteration, but then drops in later iterations. This suggests that current LLMs cannot accurately decide, under an incorrect implementation, whether the generated specification has already captured the requirement and should stop being revised; continued repair may instead move the specification toward the faulty implementation or other non-faithful forms.

\noindent\textbf{Answer to RQ3.}
\textsc{RULER} (2-stage) does not consistently improve LLM-generated specifications. In contrast, \textsc{RULER} (2-stage+refine) increases the final success rate from 18.47\% to 24.27\% on correct implementations and from 24.95\% to 29.49\% on incorrect implementations. However, refinement is less stable on incorrect implementations: GPT-5.5 performs worse after refinement than with \textsc{RULER} (2-stage). This result suggests that iterative feedback helps repair specifications when the implementation is correct, but LLMs still have difficulty distinguishing requirement-consistent specifications from specifications that are merely aligned with faulty implementation behavior.

\begin{table*}[t]
\centering
\caption{Performance of AutoSpec and advanced prompting approaches compared with basic prompting in RQ3.}
\label{tab:rq3-advanced-overall}
\scriptsize
\setlength{\tabcolsep}{3pt}
\renewcommand{\arraystretch}{1.05}
\resizebox{0.9\textwidth}{!}{
\begin{tabular}{llccccccc}
\toprule
\multirow{2}{*}{Strategy} & \multirow{2}{*}{Model} &
\multicolumn{3}{c}{Correct implementations} & &
\multicolumn{3}{c}{Incorrect implementations} \\
\cline{3-5}\cline{7-9}
& & Syn. & Success & Success|Syn. & & Syn. & Success & Success|Syn. \\
\midrule
\multirow{3}{*}{Basic prompt}
& gpt-5.4-mini & 269 (53.80\%) & 73 (14.60\%) & 27.14\% & & 490 (54.26\%) & 222 (24.58\%) & 45.31\% \\
& gpt-5.5 & 384 (76.80\%) & 126 (25.20\%) & 32.81\% & & 727 (80.51\%) & 235 (26.02\%) & 32.32\% \\
& deepseek-v4-pro & 253 (50.60\%) & 78 (15.60\%) & 30.83\% & & 513 (56.81\%) & 219 (24.25\%) & 42.69\% \\
\midrule
\multirow{3}{*}{AutoSpec}
& gpt-5.4-mini & 207 (41.40\%) & 69 (13.80\%) & 33.33\% & & 401 (44.41\%) & 171 (18.94\%) & 42.64\% \\
& gpt-5.5 & 260 (52.00\%) & 85 (17.00\%) & 32.69\% & & 491 (54.37\%) & 225 (24.92\%) & 45.82\% \\
& deepseek-v4-pro & 225 (45.00\%) & 78 (15.60\%) & 34.67\% & & 441 (48.84\%) & 186 (20.60\%) & 42.18\% \\
\midrule
\multirow{3}{*}{\textsc{RULER} (2-stage)}
& gpt-5.4-mini & 262 (52.40\%) & 66 (13.20\%) & 25.19\% & & 492 (54.49\%) & 220 (24.36\%) & 44.72\% \\
& gpt-5.5 & 395 (79.00\%) & 110 (22.00\%) & 27.85\% & & 727 (80.51\%) & 256 (28.35\%) & 35.21\% \\
& deepseek-v4-pro & 243 (48.60\%) & 72 (14.40\%) & 29.63\% & & 476 (52.71\%) & 204 (22.59\%) & 42.86\% \\
\midrule
\multirow{3}{*}{\textsc{RULER} (2-stage+refine)}
& gpt-5.4-mini & 356 (71.20\%) & 88 (17.60\%) & 24.72\% & & 676 (74.86\%) & 291 (32.23\%) & 43.05\% \\
& gpt-5.5 & 472 (94.40\%) & 166 (33.20\%) & 35.17\% & & 846 (93.69\%) & 251 (27.80\%) & 29.67\% \\
& deepseek-v4-pro & 369 (73.80\%) & 110 (22.00\%) & 29.81\% & & 683 (75.64\%) & 257 (28.46\%) & 37.63\% \\
\bottomrule
\end{tabular}
}
\end{table*}

\subsection{RQ4: comparison with non-formal approaches}

Due to the extremely costly nature of HoarePrompt, in RQ4, we only conduct our comparison on 200 problems with a total of 557 implementations, among which 200 are correct and 357 are incorrect.
Like previous RQs, we also perform E-ACSL-based classification: the ``correct'' class are treated as the correct prediction, and other classes except ``other'' are treated as incorrect prediction.
We follow the original HoarePrompt paper and report the balanced accuracy (BA), true positive rate (TPR), false positive rate (FPR), and Matthews correlation coefficient (MCC).

\begin{table*}[t]
\centering
\caption{Comparison between the non-formal approach of HoarePrompt and our formal specification generation approach on predicting the consistency between requirements and code. ``All samples'' reports results on all 557 samples, while ``Syn.-correct'' reports results after excluding syntactically invalid LLM-generated ACSL specifications.}
\label{tab:rq4-nonformal}
\scriptsize
\setlength{\tabcolsep}{3pt}
\renewcommand{\arraystretch}{1.05}
\resizebox{0.9\textwidth}{!}{
\begin{tabular}{llrrrrcrrrrr}
\toprule
\multirow{2}{*}{Approach} & \multirow{2}{*}{Model} &
\multicolumn{4}{c}{All samples} &
\multicolumn{1}{c}{} &
\multicolumn{4}{c}{Syn.-correct} &
\multirow{2}{*}{Avg. Tokens} \\
\cline{3-6}\cline{8-11}
& & BA & TPR & FPR & MCC & & BA & TPR & FPR & MCC & \\
\midrule
\multirow{3}{*}{HoarePrompt} & gpt-5.4-mini & 0.7017 & 0.6050 & 0.2017 & 0.4066 & & 0.7082 & 0.6280 & 0.2115 & 0.4156 & 45,918 \\
& gpt-5.5 & 0.9053 & 0.8750 & 0.0644 & 0.8124 & & 0.9139 & 0.8820 & 0.0543 & 0.8315 & 56,995 \\
& deepseek-v4-pro & 0.7679 & 0.7850 & 0.2493 & 0.5175 & & 0.7836 & 0.8120 & 0.2447 & 0.5464 & 58,767 \\
\midrule
\multirow{3}{*}{HoarePrompt + unroll} & gpt-5.4-mini & 0.6968 & 0.8250 & 0.4314 & 0.3817 & & 0.6968 & 0.8250 & 0.4314 & 0.3817 & 146,511 \\
& gpt-5.5 & 0.7127 & 0.7700 & 0.3445 & 0.4082 & & 0.7127 & 0.7700 & 0.3445 & 0.4082 & 216,359 \\
& deepseek-v4-pro & 0.6373 & 0.6500 & 0.3754 & 0.2639 & & 0.6373 & 0.6500 & 0.3754 & 0.2639 & 139,401 \\
\midrule
\multirow{3}{*}{Basic prompt} & gpt-5.4-mini & 0.1959 & 0.1460 & 0.7542 & -0.5842 & & 0.3622 & 0.2714 & 0.5469 & -0.2650 & 7,346 \\
& gpt-5.5 & 0.2561 & 0.2520 & 0.7398 & -0.4713 & & 0.3257 & 0.3281 & 0.6768 & -0.3337 & 12,924 \\
& deepseek-v4-pro & 0.1993 & 0.1560 & 0.7575 & -0.5783 & & 0.3676 & 0.3083 & 0.5731 & -0.2492 & 7,565 \\
\midrule
\multirow{3}{*}{\textsc{RULER} (2-stage)} & gpt-5.4-mini & 0.1878 & 0.1320 & 0.7564 & -0.5994 & & 0.3495 & 0.2519 & 0.5528 & -0.2881 & 11,887 \\
& gpt-5.5 & 0.2517 & 0.2200 & 0.7165 & -0.4771 & & 0.3153 & 0.2785 & 0.6479 & -0.3531 & 20,299 \\
& deepseek-v4-pro & 0.1850 & 0.1440 & 0.7741 & -0.6065 & & 0.3624 & 0.2963 & 0.5714 & -0.2605 & 16,061 \\
\midrule
\multirow{3}{*}{\textsc{RULER} (2-stage+refine)} & gpt-5.4-mini & 0.2491 & 0.1760 & 0.6777 & -0.4806 & & 0.3388 & 0.2472 & 0.5695 & -0.3075 & 52,551 \\
& gpt-5.5 & 0.3050 & 0.3320 & 0.7220 & -0.3789 & & 0.3242 & 0.3517 & 0.7033 & -0.3413 & 110,565 \\
& deepseek-v4-pro & 0.2523 & 0.2200 & 0.7154 & -0.4760 & & 0.3372 & 0.2981 & 0.6237 & -0.3108 & 65,123 \\
\bottomrule
\end{tabular}
}
\end{table*}

Table~\ref{tab:rq4-nonformal} shows that current LLM-based formal specification generation is still not effective for predicting the consistency between natural-language requirements and code. HoarePrompt obtains much stronger classification performance across models: standard HoarePrompt reaches an average BA of 0.7916 and an average MCC of 0.5788 on all samples, while the best formal-specification variant, \emph{Ours (\textsc{RULER} (2-stage+refine))}, reaches only 0.2688 BA and -0.4452 MCC on average. Although our basic prompt and \textsc{RULER} (2-stage) consume fewer tokens than HoarePrompt, the lower cost does not translate into useful classification performance.

The main reason is the low true-positive rate of the formal-specification pipeline. In this setting, a correct implementation is predicted as correct only when the generated ACSL specification can be verified. However, our variants achieve much lower TPR than HoarePrompt across all models, indicating that many correct implementations are rejected because the generated specifications are syntactically invalid, unverifiable, or not faithful to the requirement. Filtering out syntactically invalid specifications improves the results, but the MCC values remain negative. This suggests that syntax errors explain only part of the gap; the deeper limitation is that LLMs do not reliably generate formal specifications that capture the intended requirement.

The comparison with HoarePrompt + unroll also shows that more reasoning steps are not necessarily beneficial. Unrolling substantially increases token consumption and often increases the false-positive rate, leading to lower MCC than standard HoarePrompt for all three models. Overall, RQ4 suggests that, under the current generation quality, LLM-based formal specification generation is less reliable than non-formal LLM reasoning for judging requirement-code consistency. Formal verification can provide stronger evidence only when the generated specification is faithful, which remains the main bottleneck.

\noindent\textbf{Answer to RQ4.}
Compared with HoarePrompt, current LLM-based formal specification generation cannot yet effectively predict requirement-code consistency. Its main limitation is the low true-positive rate: LLMs often fail to generate ACSL specifications that are both verifiable and faithful to the natural-language requirement, causing many correct implementations to be classified as inconsistent.

\section{Threats to Validity}

\textbf{Internal validity.}
An internal threat comes from the preprocessing of CodeContests programs. We rewrite the original input/output logic into function arguments and return values to make the programs compatible with ACSL and Frama-C. Although we validate the rewritten programs against the provided test cases, these tests may not cover all behaviors, and some semantic changes may remain undetected. Another threat comes from LLM nondeterminism: different runs may produce different specifications even under the same prompt. To mitigate this threat, we use a fixed experimental setup for all compared settings and apply the same verification pipeline to all generated specifications.

\textbf{Construct validity.}
Our metrics approximate specification quality through syntactic correctness, Frama-C verification outcomes, and manual classification. These measures do not fully capture semantic faithfulness to the natural language requirement. In particular, verification failure on an incorrect implementation may indicate a faithful specification, but it may also result from an inconsistent, overly strong, or ill-formed specification. Conversely, verification success does not always imply a requirement-faithful specification, because the specification may be weak or misled by the implementation. We therefore complement verification-based metrics with manual inspection categories, including correct, wrong, weak, and misled specifications.

\textbf{External validity.}
Our benchmark is built from programming contest problems, whose requirements are usually more detailed and less ambiguous than industrial software requirements. The programs are also relatively small compared with real-world systems. As a result, the findings may not directly generalize to large codebases, domain-specific APIs, or requirements that involve complex environmental assumptions. Nevertheless, given that formal verification on real-world software repositories is still extremely difficult and costly, programming contests provide a controlled setting with natural language requirements and both correct and incorrect implementations, making them suitable for studying the core bimodal specification generation problem.


\section{Conclusion}

We conduct the first systematic study on LLM-based formal specification generation from both natural-language requirements and source code, where the implementation may be either correct or incorrect. The evaluation results show that bimodal specification generation is substantially harder than generating specifications from correct code alone. When implementations are correct, adding requirements does not consistently improve verification success over code-only prompting; when implementations are incorrect, LLMs may generate specifications that are weak, misled by the implementation, or fail to expose the mismatch between the requirement and the code, and these limitations cannot be fully eliminated by advanced prompting techniques. Compared with the non-formal HoarePrompt approach, LLM-generated ACSL specifications are less effective in predicting the consistency between requirements and implementations. These findings suggest that LLM-based formal specification generation still has a long way to go before it can reliably verify programs against their requirements. Future work in LLM-based formal specification generation should focus more on requirement-faithful specification generation, and tighter integration between LLM reasoning and deductive verification feedback.

\bibliographystyle{ACM-Reference-Format}
\bibliography{IEEEtranBST2/IEEEexample}

\end{document}